\documentclass[aps,prl,twocolumn,showpacs,final]{revtex4-1}
\usepackage{graphicx}
\usepackage{color}
\usepackage{marvosym }
\usepackage{kantlipsum}
\usepackage{dsfont}
\usepackage{physics}
\usepackage{xcolor}
\usepackage{breqn}
\usepackage{amsmath}
\usepackage{hyperref}
\usepackage{amssymb}
\usepackage{amsfonts}
\usepackage{wrapfig}
\usepackage{multirow}
\usepackage{bm}
\usepackage{dcolumn}% Align table columns on decimal point
\newcommand{\etal}{\textit{et al.\ }}
\newcommand{\ie}{\textit{i.e.\ }}
\newcommand{\eg}{\textit{e.g.\ }}
\newcommand{\etc}{\textit{etc.\ }}
\newtheorem{prop}{Proposition}[section]
\newcommand\at[2]{\left.#1\right|_{#2}}
\newcommand\pN{\mathcal{N}}
\begingroup\makeatletter
\catcode`,=\active
\global\let\breqn@comma,
\protected\gdef,{\ifmmode\expandafter\breqn@comma\else\expandafter\active@comma\fi}
\endgroup

\begin{document}
\title{Information flow in political elections: a stochastic perspective}
\author{\href{http://santoshkumarradha.me }{Santosh Kumar Radha}}
\affiliation{Department of Physics, Case Western Reserve University, 10900 Euclid Avenue, Cleveland, OH-44106-7079}
\begin{abstract}
	Often times, a candidate's attractiveness is directly associated with his clear ideologies and opinions on various policies and social issues. Using the ideas of stochastic differential equations and Ornstein-Uhlenbeck Process, we develop a phenomenological model to understand the effect of (un)clearly communicating a candidate's stance on policies to the voting public. We will show that, counter intuitively, there are quantifiable advantages to be vague on one's stance. 
	% We also try to answer the question - When is the right time, from the start to the election day, for a candidate to announce his \textit{unpopular} stance on policies.
\end{abstract}
\maketitle
Quantitative analysis of political elections have been a staple for many years. Models of election are many and varied, each with it’s own focus. There is a vast literature on methods to forecast the elections using tools like fundamental indicators\cite{hummel2013fundamental}, market indicators\cite{berg2008results}, Bayesian methods\cite{linzer2013dynamic} and even social media strategies \cite{gayo2012no} and other general statistical tools \cite{hummel2014fundamental,klarner2008forecasting,lauderdale2015under}. Although there is extensive work on modeling of political dynamics between candidates \cite{bottcher2018clout,braha2017voting,fernandez2014voter,galam2004dynamics,radha2019stochastic}, most of the study focus on global/macro perspective while either neglecting or including the complex effects of individual components (\eg stances on different policies) as a \textit{mean field} effect.

In this letter, we develop a quantitative phenomenological model to understand and analyze the effect of one such individual component - efficient information flow in an election/voting system. Information flow is a fundamental notion in many areas from physics to quantitative finance. Whenever a decision is being made on a candidate, one typically encounters uncertainties about the candidate's stance on various policies. These uncertainties can occur due to various reasons from  candidate's lack of effective communication to spread of  deliberate misinformation. While qualitatively, inefficient transfer of information from candidate to public might seem like a negative effect, we show that there are certain cases where these inefficiencies are not only advantages, but also necessary.  We use Ornstein-Uhlenbeck equations\cite{PhysRev.36.823} to model the time dependent information flow between the candidate and public. Use of stochastic theory to model complex elections have also been used by Fenner\etal \cite{fenner2018stochastic} to analyze the polls leading up to the UK 2016 EU referendum. We will start by introducing the model, after which we will explore the properties and effects of various parameters phenomenologically.

% \kant[2]
% \kant[3]

\paragraph*{Model}: Let us denote the stance of the candidate on policy $p$ as $\mu$ with his own uncertainty on the stance being $\sigma$, where $-\infty<\mu<\infty$ with positive values being in favor for policy and negative values being against. This could model a variety of situations from social issues to ideologies. For example, in case of left vs right, a left leaning moderate candidate might have $\mu\approx 0.5,\sigma\approx 0.5$ with $\mu>0$ and $\mu<0$ being left and right respectively. Now we model the public \textit{perceived} stance of the candidate's policy at time $t$ during the election to by a random variable  $X_t$. At the start of election ($t=0$) (or the start of announcing the candidacy), $\at{X_t}{t=0} = X_0=\pN(\mu_0,\sigma_0)$ \ie at $t=0$, the public has a predisposed idea on what the candidate's stance is with a distribution given by $X_0$. For instance, $X_0=\mu_0=0$ would mean that the public has no idea of what the candidate's stance is. As time flows, the candidate's opinion on $p$ is made clearer though various information transfer processes like public speech, social media posts \etc and $\at{E\left[X_t}{t\rightarrow\infty}\right]\rightarrow \mu$. This system is modeled as a solution to a stochastic differential equation (SDE) given by 
\begin{equation}
\mathrm{d} X_{t}=r\left(\mu-X_{t}\right) \mathrm{d} t+\sigma \mathrm{d} W_{t}, \quad t>0 \label{eq:1}. 
\end{equation}

where $r$ is the rate at which the candidate makes his stance $\mu$ clear with a variance of $\sigma\geq0$. $W_{t}$ is a standard Brownian Motion on $\mathrm{R}$. \autoref{eq:1} can alternatively be written in terms of stochastic integral form as 
\begin{equation}
X_{t}=\mu\left(1-e^{-r t}\right)+\sigma e^{-r t} \int_{0}^{t} e^{r s} \mathrm{d} W_{s}+X_{0} e^{-r t}, \quad t \geq 0 \label{eq:2}
\end{equation}

It is trivial to check that \eqref{eq:2} is the unique, strong Markov sollution to \eqref{eq:1}\cite{protter2005stochastic}. This is Ornstein-Uhlenbeck Process, which is an extension of Brownian motion with friction. The important feature of \eqref{eq:1}, is that the expectation value at time $t$ is given by

\begin{equation}
E \left[X_{t}\right]=\mu\left(1-e^{-r t}\right)+e^{-r t} E \left[X_{0}\right],\quad t \geq 0\label{eq:3},
\end{equation}
which gives us the required asymptotic Gaussian behavior $\at{E\left[X_t}{t\rightarrow\infty}\right]\rightarrow \mu$. These type of processes are often refereed to as \textit{mean-reverting} processes. Although there are different generalizations of Ornstein-Uhlenbeck Process \cite{,dixit1994investment}, we here choose the vanilla model to introduce the phenomenology. 

In general, the formal solution of \eqref{eq:1} is given by $X_t=\pN(E \left[X_{t}\right],\operatorname{Var}[X_t])$ where
\begin{align}
\operatorname{Var}[X_t]=
\frac{\sigma^{2}}{2 r}\left(1-e^{-2 r t}\right)+e^{-rt} \operatorname{Var} \left[X_{0}\right]\label{eq:4}
\end{align}

\begin{figure}[htbp]
\includegraphics[width=\columnwidth]{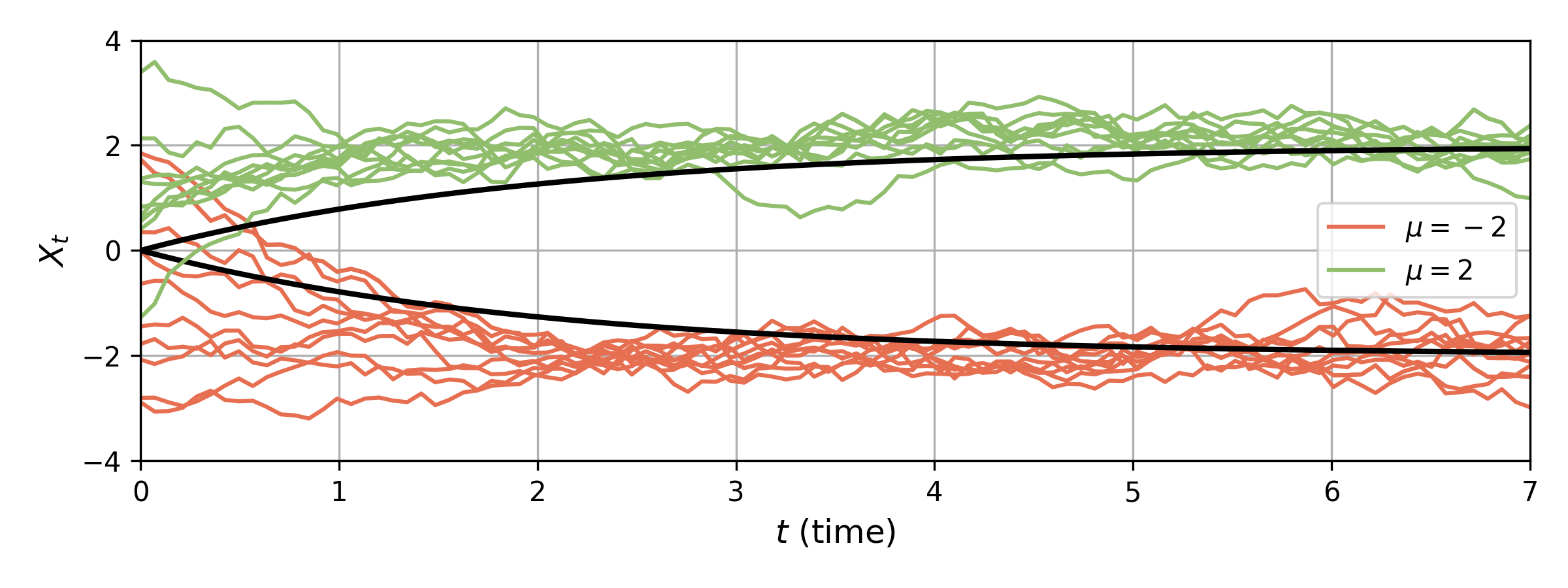}
\caption{Sample trajectories of $X_t$ starting from $X_0\approx\pN(0,2)$ for $\sigma=0.5$, $r=1$, (green) $\mu=2$ and (red) $\mu=-2$ }
\label{fig:path}
\end{figure}

\ref{fig:path} shows a numerical simulation of $X_t$ for values of $\mu=2$ (positive stance) and $\mu=-2$ (negative stance) for 10 different stochastic curves starting with an initial distribution of $\pN(0,2)$ in green and red respectively. Black lines show the evolution of mean according to \eqref{eq:3}. As one can see, initial idea of candidate's perspective on policy is quite spread out which then converges to $\mu$ as time proceeds. 

In this model, as seen from \eqref{eq:4}, $\operatorname{Var}[X_t]$ plays the vital role of how collectively \textit{confused} the public is about the candidate's opinion at any given time $t$. This is mainly governed by the variable $\frac{\sigma}{2r}$. It is interesting to note that this confusion is not dependent on how strong a negative or positive stance the candidate takes \ie $\mu$. 

With these notion in mind, we now propose that the probability that the candidate wins ($P_T$)

\begin{prop}
\label{prop1}
Probability  ($P_T$) that the candidate wins maximum votes when the election is at time $T$ is given by Hellinger distance measure $1-H(X_T,\tilde{X}_p)$ where  $\tilde{X}$ is the public collective public opinion on policy $p$.
\end{prop}

Hellinger distance $H$ is given by\cite{hellinger1909neue} 
\begin{equation}
H^{2}(f, g)=
\frac{1}{2} \int(\sqrt{f(t)}-\sqrt{g(t)})^{2} d t\label{eq:5}
\end{equation}
where $f,g$ are two continuous probability density functions. Because Hellinger distance is a bounded metric on the space of probability distributions, we can directly relate it to the probability of winning. Prop.\ref{prop1} is nothing but a measure of how close the candidate's perceived opinion aligns with actual public onion at any given time, matching of which would dictate the winning.

Based on \eqref{eq:3} and \ref{eq:4}, we have $X_T$ to be a Gaussian density function and if the public stance on $p$ can be modeled as Gaussian, \eqref{eq:5} reduces to 
\begin{equation}
P_T=\sqrt{\frac{2 \sigma_{T} \tilde{\sigma}}{\sigma_{T}^{2}+\tilde{\sigma}^{2}}} \exp{-\frac{1}{4} \frac{\left(\mu_{T}-\tilde{\mu}\right)^{2}}{\sigma_{T}^{2}+\tilde{\sigma}^{2}}}\label{eq:6},
\end{equation}
where $\sigma_T,\mu_T$ are the variance and mean of the random variable $X_T$ and public stance $=\pN(\tilde{\mu},\tilde{\sigma})$. Using \eqref{eq:6}, we can now study the effect of various parameters in the model and its result on the probability of winning.

Based on \eqref{eq:6}, as a sanity check, we will first explore the effect of candidates controversial take on policy $p$, \ie $\abs*{\mu-\tilde{\mu}}>>0$. Since $\sigma_T$ is unaffected by $\mu$, we will assign $\alpha_T=2 \sigma_{T} \tilde{\sigma}$ and $\beta_T=\sigma_{T}^{2}+\tilde{\sigma}^{2}$, then 
\begin{dmath}
P_T(\mu)=\sqrt{\frac{2\alpha_T}{\beta_T}}\exp{\frac{-1}{4\beta_T}\left(\mu\left(1-e^{-r T}\right)+e^{-r T} E \left[X_{0}\right]-\tilde{\mu}\right)^2}\label{eq:7},
\end{dmath}

Which as $T\rightarrow \infty$ is a Gaussian distribution with mean $\mu-\tilde{\mu}$, which essentially says that when every other variable $=1$, the maximum likelihood of winning the election is when the public and candidate have the same stance on $p$.

\begin{figure}[htbp]
\includegraphics[width=\columnwidth]{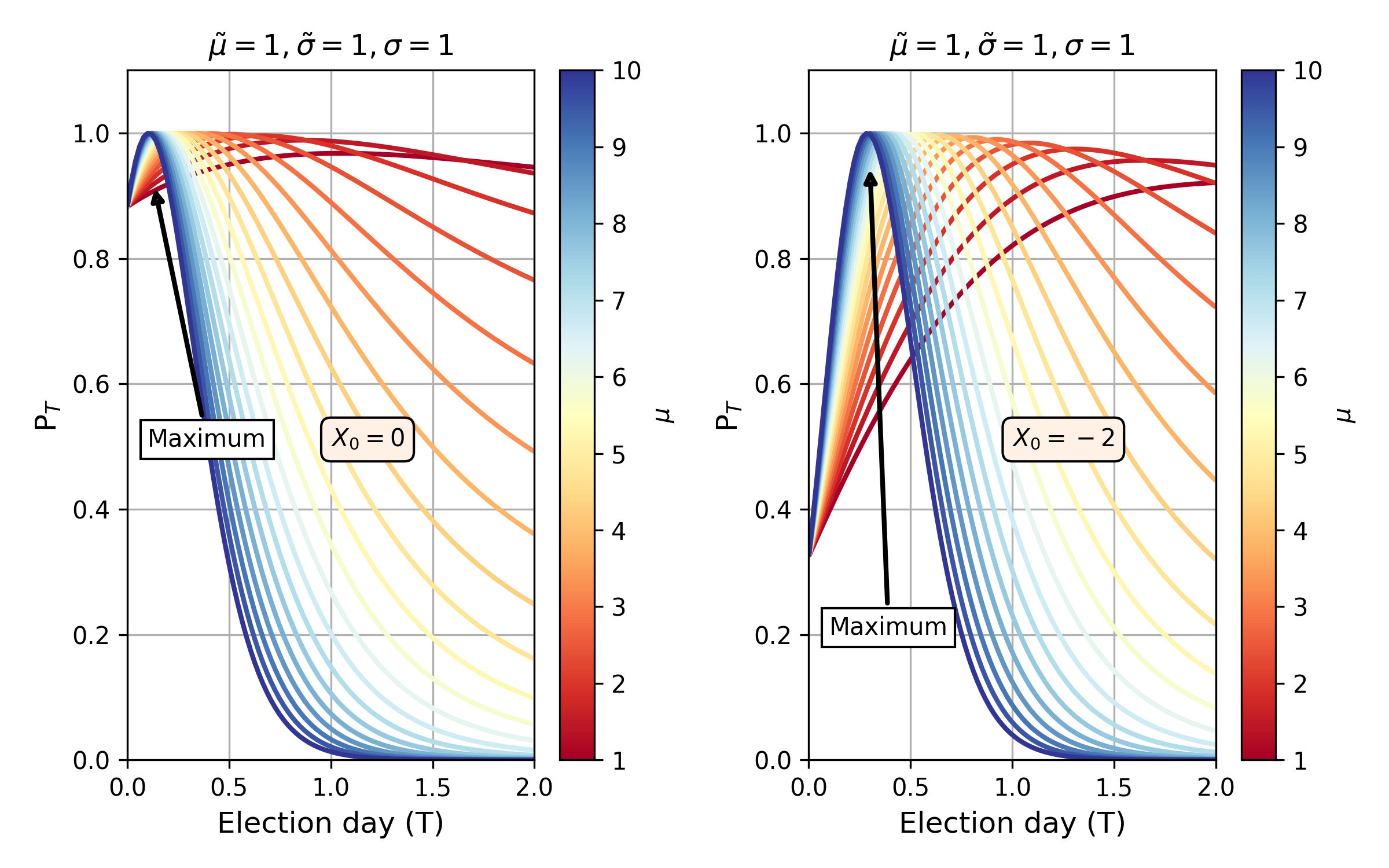}
\caption{ $P_T$ for $0<T<2$ when $\tilde{\mu}=1,\tilde{\sigma}=1,\sigma=1$ and $1<\mu<10$ for starting distribution with (a) $X_0=0$ (b) $X_0=-2$ }
\label{fig:pt_mu}
\end{figure}

Things get more interesting if we study the likelihood of winning as a function of number of days to election. \autoref{fig:pt_mu} shows the evolution of $P_T$ as a function of number of days from election for all quantities set to 1, except $\mu$, which varies from $1$ to $10$. As one can clearly see from (a), for $\tilde{\mu}=1$ (the majority accepted stance on policy = 1) and the candidate's stance is far away from it $\mu=10$, at $T\rightarrow\infty$, the candidate has $\approx0$ chance of winning, as mentioned in previous paragraph. But, when one closely observes the figure, we see that there is an inflation point where the candidate has the maximum likelihood to win. This confirms a very intuitive strategy that, when you have an unpopular opinionated candidate/policy, he is more likely to win when announcing candidacy/stance very close to election as the public cannot yet fully digest his actual stance despite being clear due to lack of time to assimilate. In (b), we start from an already negative perception on the candidates stance \ie $X_0=-2$, and this inflation point always survives.

\begin{figure}[htbp]
\includegraphics[width=\columnwidth]{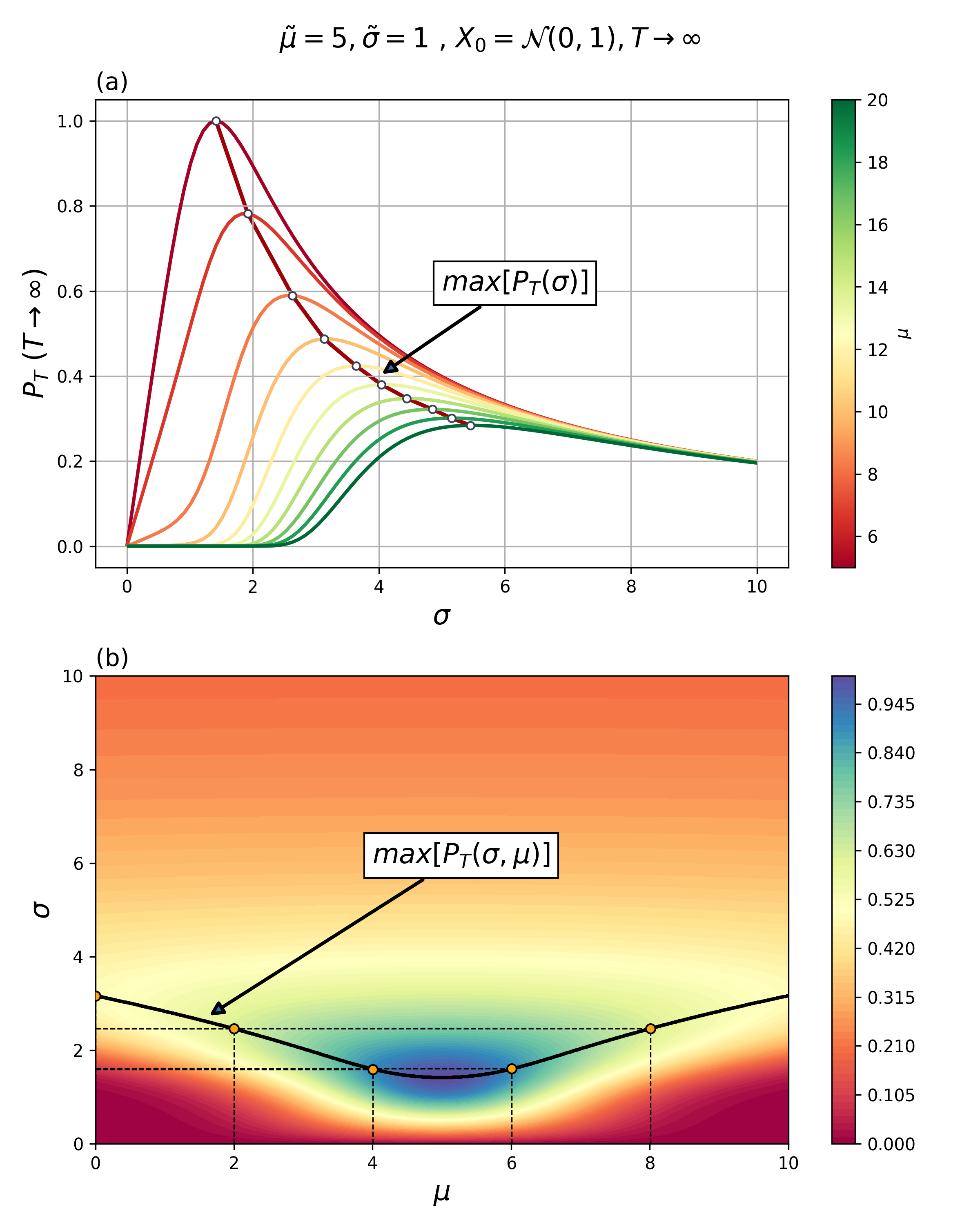}
\caption{ $P_T$ for $0<T<2$ when $\tilde{\mu}=1,\tilde{\sigma}=1,\sigma=1$ and $1<\mu<10$ for starting distribution with (a) $X_0=0$ (b) $X_0=-2$ }
\label{fig:mu-sig}
\end{figure}

Now, we focus on the effect of introducing \textit{confusion} \ie $\sigma$, for a given stance. We will focus on $T\rightarrow\infty$ limit for clarity as extending to finite $T$ is straightforward. \autoref{fig:mu-sig} (a) shows the calculated $P_T$ for various $\mu$ as a function of $\sigma$, with $\tilde{\mu}=5$ and $\tilde{\sigma}=1$, since we are taking $T\rightarrow\infty$, one should note that the initial starting point ($X_0$) does not matter as we are interested in the asymptotic behavior.

We start with $\mu=5$ (red curve), where we have the public stance given by $\tilde{\mathcal{N}}(5,1)$, which has the maximum probability reaching when $\sigma=1$, giving us that maximum likelihood is reached when  $\at{X_T}{T\rightarrow\infty}=\tilde{\mathcal{N}}(5,1)$. As the candidate's stance differs from popular stance \ie $\mu\neq\tilde{\mu}$ (red to green curve), his probability to win reduces. Despite this reduction, one can still tune $\sigma$ to reach the optimal probability. This essentially dictates that when an unpopular stance is held by a candidate, there is an optimal noise that can be added while communicating to public which maximizes the chances of winning. From (a), one can also see that as $\mu-\tilde{\mu}\rightarrow\infty$, the effect of optimizing $\sigma$ does not create appreciable advantage. Thus, when a polarizing candidate is introduced, the best bet is to be extremely confusing about the stance and very little is to be gained by optimizing $\sigma$. 

\paragraph*{Polar opinions}: One important use case of the above formulation is when there is a binary stance by the public on a policy. For instance, there can be scenarios where the public opinion on a policy $p$ is given by $\tilde{x} \in \{-1,1\}$ while the candidate can have his stance $x \in (-\infty,\infty)$. We now have the random variable $X\in (-\infty,\infty)$, which the public perceive as the candidate's stance, based on which at election time $(T)$, public makes the choice $\tilde{X}_T$ based on candidate's perceived stance $X_T$, given by

\begin{equation}
    \tilde{X}_T = \begin{cases}
              +1 & \text{if } X_T> 0,\\
              -1 & \text{if } X_T< 0.\\
          \end{cases}\label{eq:8}
\end{equation}

\autoref{eq:8} says that, at the election date, the public makes the decision based on the candidate's  positive/negative stance on the policy, irrespective how how strong or weak the stance is. Following the phonological model, the important quantity to calculate is the probability kernel of \eqref{eq:1}, which is given by 

\begin{dmath}
\mathcal{P}(t, x, y)=\frac{1}{\sqrt{\pi \sigma^{2}\left(1-e^{-2 r t}\right) / r}} \exp {-\frac{\left(y-\mu-(x-\mu) e^{-r t}\right)^{2}}{\sigma^{2}\left(1-e^{-2 r t}\right) / r} }\label{eq:9},
\end{dmath}

where $\mathcal{P}$ is the probability kernel for $X$ to reach $y$ from $x$ in time $t$ with $\mu,\sigma$ being the mean and spread of candidate's stance. Using \eqref{eq:9}, one can calculate the ideal time to start the election\footnote{Time to start the election is the same as time till the election after announcing candidacy}. We will illustrate this with an example where we know that the majority favors $+1$ \ie $\tilde{\mu}=+1$ and $\tilde{\sigma}=0$ (this information can be extracted from public polls \etc). In this case, we first need to calculate the probability ($P^\dagger(t,x)$) for a point $x$ to reach $y>0$ in a time $t$, 
\begin{dmath}
	P^\dagger(t,x)=\int_0^\infty \mathcal{P}(t,x,y)dy= 1 - \frac{1}{2}\operatorname{erfc}{\left(- \frac{\sqrt{r} \left(- \mu - \left(- \mu + x\right) e^{- r t}\right)}{\sigma \sqrt{1 - e^{- 2 r t}}} \right)}\label{eq:10}
\end{dmath}

\begin{figure}[!h]
\includegraphics[width=\columnwidth]{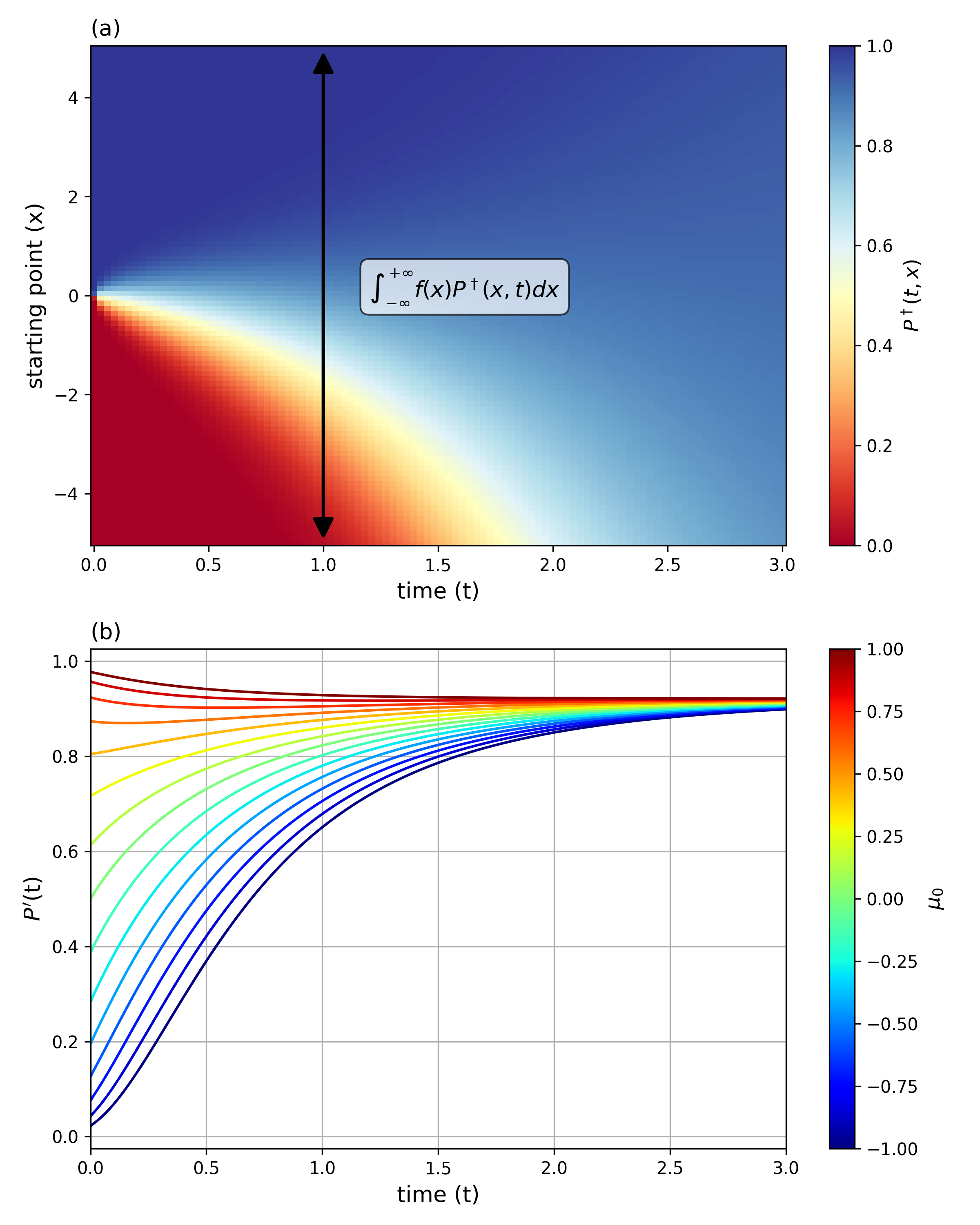}
\caption{ $P_T$ for $0<T<2$ when $\tilde{\mu}=1,\tilde{\sigma}=1,\sigma=1$ and $1<\mu<10$ for starting distribution with (a) $X_0=0$ (b) $X_0=-2$ }
\label{fig:pt}
\end{figure}

\autoref{fig:pt}(a) shows $P^\dagger(t,x)$ for $\mu=\sigma=r=1$. This is the probability that a person who thinks that the candidate's stance is $x$ would choose $\tilde{X}_t=1$, qt time $t$. Thus under the assumption that majority favors $+1$, this would guarantee that the person is voting for the candidate. Using $P^\dagger(t,x)$, we can now calculate exactly how long to wait for the election for various starting points. For example, if we know that at this given moment $t=0$, the public perceives the stance of candidate to be $f(x)$ where $f(x):=\pN(\mu_{0},\sigma_0)$, then the probability that they reach $+1$ at time $t$ is given by

\begin{dmath}
P^\prime(t)=\int_{-\infty}^{+\infty} f(x) P^\dagger(t,x) dx. \label{eq:11}
\end{dmath}

This is easy to \autoref{fig:pt}(a), as we integrate out $x$, sampling from the points where we start \ie  $\pN(\mu_{0},\sigma_0)$ (shown in black line (a)). We show $P^\prime (t)$ in (b) for various starting point $\mu_0$. It is clear that when the public perceives the candidate to be on the negative side, there is a huge advantage to wait (assuming the public stance on policy is positive $\tilde{\mu}=+1$ ) while the inverse is true if they perceive him to be on the positive side. Although we discussed the case for majority favors $+1$, this can easily be extended to $-1$.

Finally, \eqref{eq:1} can be used as a base to model and study other intricate effects. For instance one can study effect of sudden impact full negative/positive news by simply adding terms to \eqref{eq:1} to include jump process which can be analyzed numerically. Recent work by Brody \etal\cite{brody2018model} showed that \textit{Fake news} can be modeled by adding a noise with non-zero drift.

%------------------------------

Even though all previous discussions pertained to single policy, one can easily generalize this model to accommodate multiple policy by increasing the dimensionality of the random variable. This is important because public decide between candidates based on number of different policies that are both quantifiable and non quantifiable. Subconsciously, these stances are mapped into an overall score, and the public at the end votes for the candidate with the highest score. Similar to \eqref{eq:1}, one can define the $n-$dimensional OU process with $\mathbf{X_t}=(X^1_t,X^2_t,\ldots,X^n_t)$ which are random variables for policies $\mathbf{p}=(p^1,p^2,\ldots,p^n)$ satisfying

\begin{equation}
\mathrm{d} \mathbf{X}^i_{t}=\sum_k R^{ik}\left(M^k-\mathbf{X}^k_{t}\right) \mathrm{d} t+\sum_l \Sigma^{il} \mathrm{d} \mathbf{W}^l_{t}, \quad t>0 \label{eq:8},
\end{equation}
where $R$ and $\Sigma$ are $n\times n$ matrices and $M$ is a $n \times 1$ matrix while $\mathbf{W_{t}}$ is a vector of $n$ independent Brownian motions and repeated. Formal solution of \eqref{eq:8} is given by time dependent Gaussian vector\cite{vatiwutipong2019alternative}

\paragraph*{Summery}: In this work, we have developed a quantitative phenomenological model to understand the effect of uncertainty in information flow in elections. This offers a initial model upon which complexities can be adorned to quantitatively analyze the effects of isolated variables, which helps in quantitative strategizing of election campaigns. Further, we showed the existence of situations where ineffective information flow can be advantageous. Finally, we noted that this formalism, although developed for single policy study, can be extended to multiple policies and is still analytically tractable.
\bibliography{lmto}
\end{document}